\documentclass[aps, twocolumn, amsmath, amssymb, prb, 10pt, floatfix]{revtex4-2}

\newif\ifincludefigures
\includefigurestrue 

\ifincludefigures\fi

\usepackage[per-mode=symbol,separate-uncertainty]{siunitx}

\usepackage[]{graphics} 
\usepackage{amsmath,color,colordvi}  
\usepackage{mathtools}
\usepackage[pdftex]{graphicx} 
\usepackage[english]{babel}   
\usepackage{braket}

\definecolor{navyblue}{rgb}{0.0, 0.0, 0.5}
\usepackage[protrusion=true, expansion=true]{microtype} 
\usepackage[colorlinks=true, breaklinks=false, linkcolor=navyblue, urlcolor=navyblue, citecolor=navyblue]{hyperref}



\makeatletter
\renewcommand{\figurename}{\textbf{Figure}}
\renewcommand{\tablename}{\textbf{Table}}

\renewcommand{\fnum@figure}{\textbf{\figurename~\thefigure}}
\renewcommand{\fnum@table}{\textbf{\tablename~\thetable}}
\makeatother

\begin{document}

\title{Initial demonstration of a quantum heat engine based on dissipation-engineered superconducting circuits}

\author{Tuomas Uusn\"{a}kki$^1$}
\email{tuomas.uusnakki@aalto.fi}
\author{Timm Mörstedt$^1$}
\author{Wallace Teixeira$^1$}
\author{Miika Rasola$^1$}
\author{Mikko M\"{o}tt\"{o}nen$^{1,2}$}
\email{mikko.mottonen@aalto.fi}

\affiliation{$^1$QCD Labs, QTF Center of Excellence, Department of Applied Physics, Aalto University, P.O. Box 13500, FI-00076 Aalto, Finland\\
  $^2$VTT Technical Research Centre of Finland, QTF Center of Excellence, P.O. Box 1000, FI-02044 VTT, Finland}
 
\date{\today}

\begin{abstract}
Quantum heat engines require precise control over thermal reservoirs and the energies of the quantum working medium. Although superconducting circuits enable accurate engineering of controlled quantum systems, they have not yet been employed to experimentally realize a cyclic quantum heat engine. Here, we demonstrate a quantum heat engine with superconducting circuits, using a quantum-circuit refrigerator as a tunable heat reservoir and a flux-tunable transmon qubit as the working medium. Starting from a thermal state, we implement a few quantum Otto cycles with a tailored reservoir drive inducing sequential cooling and heating, interleaved with flux ramps controlling qubit frequency. Utilizing single-shot qubit readout, we monitor the qubit state evolution during the cycles and measure positive output powers and efficiencies, agreeing with corresponding simulations. Our results verify thermodynamic models of quantum heat engines, advance control of thermal environments, and open avenues for exploring possible quantum advantages.
\end{abstract}

\maketitle

\section*{Introduction}
Heat engines are universal devices that can convert heat flow into usable energy or work. These devices are well-established and explored in classical thermodynamics and provide practical applications throughout industries. Intriguingly, quantum systems can provide features, such as quantum superposition and interference, that may yield novel insight into the concepts of heat and work. This has been a driving force for the development of quantum thermodynamics during recent decades~\cite{gemmer_quantum_2009,kosloff_quantum_2013,deffner_quantum_2019}. 

Experimental examinations of the detailed operation of quantum heat engines can potentially showcase characteristic properties of quantum mechanics and drive the exploration of energy control in microscopic and mesoscopic quantum devices. On one hand, these quantum heat engines have been realized in multiple physical systems such as a single-particle spin coupled to single-ion motion~\cite{von_lindenfels_spin_2019,van_horne_single-atom_2020}, a trapped single-ion QHE using exceptional points~\cite{zhang_dynamical_2022}, a nuclear spin system with nuclear magnetic resonance~\cite{peterson_experimental_2019,de_assis_efficiency_2019}, a nitrogen-vacancy center in a diamond~\cite{klatzow_experimental_2019}, cold rubidium atoms with electromagnetically induced transparency~\cite{zou_quantum_2017}, and a QHE driven by atomic collisions of cesium and rubidium~\cite{bouton_quantum_2021}. On the other hand, superconducting circuits have become one of the most prominent candidates for quantum technology applications such as quantum computing~\cite{dicarlo_demonstration_2009,harrigan_quantum_2021,google_quantum_ai_and_collaborators_quantum_2025}, communication~\cite{kurpiers_deterministic_2018,axline_-demand_2018,fedorov_experimental_2021}, and sensing~\cite{gasparinetti_fast_2015,kokkoniemi_bolometer_2020,wang_quantum_2021}. Although quantum-heat-engine realizations with superconducting circuits have been proposed and studied in several recent theoretical works~\cite{niskanen_information_2007,campisi_nonequilibrium_2015,altintas_rabi_2015,karimi_otto_2016,marchegiani_self-oscillating_2016,hardal_quantum_2017,rasola_proposal_2025}, including the implementations of quantum Otto cycles, there has only been a single experimental study in superconducting quantum circuits that claims the possibility of operating a non-cyclic thermal machine in the heat engine regime~\cite{sundelin_quantum_2026}. Thus, the experimental realization of a cyclic quantum heat engine in superconducting circuits, analogous to a classical heat engine, is yet to be achieved.

Control of thermal environments in superconducting circuits has been under intense investigations in recent years covering quantum-heat-transport experiments~\cite{pekola_towards_2015,ronzani_tunable_2018}, a thermally driven quantum refrigerator~\cite{aamir_thermally_2025}, quantum measurement engines~\cite{elouard_extracting_2017,buffoni_quantum_2019,dassonneville_directly_2025}, quantum bath engineering~\cite{murch_cavity-assisted_2012,kimchi-schwartz_stabilizing_2016,sharafiev_leveraging_2025}, heat and work measurements from qubit trajectories~\cite{naghiloo_heat_2020}, and local cooling of superconducting circuits using a quantum-circuit refrigerator (QCR) as a fast in-situ dissipator~\cite{tan_quantum-circuit_2017,silveri_theory_2017,sevriuk_fast_2019,yoshioka_fast_2021,sevriuk_initial_2022,yoshioka_active_2023,viitanen_quantum-circuit_2024,teixeira_many-excitation_2024,morstedt_rapid_2025,nakamura_probing_2025}. Conveniently, the QCR can be voltage biased to effectively act as a cold or hot thermal reservoir to a coupled quantum circuit, e.g., a resonator or a qubit. Consequently, the QCR can induce desired heat flow to generate quantum thermal states, the effective temperature of which can be accurately controlled by the bias voltage and period~\cite{morstedt_rapid_2025}. Therefore, we can use this single device to achieve two-way thermal tunability, i.e., to sequentially induce heating and cooling in thermodynamic cycles without the need for multiple physical thermal reservoirs and their respective coupling circuitry. During the off-state of the QCR, we can carry out adiabatic strokes of the cycle without major interference from the dissipation since the coupling to the QCR heat reservoir is diminished. Thus, the QCR appears to be a simple device to realize proof-of-concept quantum heat engines with relatively ideal Otto cycles.

In this work, we demonstrate a quantum heat engine that realizes a quantum Otto cycle in a superconducting circuit. We present the cycle for the sample consisting of a flux-tunable transmon qubit as the working medium of the heat engine coupled to a quantum-circuit refrigerator operating as a temperature-tunable thermal reservoir. After calibration, we experimentally implement the strokes of the Otto cycle using customized driving pulses on the flux and QCR drive lines. The operation of the heat engine is demonstrated by measuring the dynamics of the transmon populations for up to three cycles of the engine using single-shot readout and a Gaussian mixture model involving four Gaussian distributions. From the temporally dependent populations, we compute the corresponding thermodynamic quantities such as mean work and absorbed heat. We observe that the heat engine exhibits positive work output and efficiency.

\section*{Results}

The thermodynamic cycle for our experiment is realized by carrying out thermodynamic strokes in an Otto scheme~\cite{deffner_quantum_2019,kosloff_quantum_2017,hardal_quantum_2017}. The heat and work terms of the cycle are derived from the first law of thermodynamics, where the internal energy of the system is defined as the expectation value of the quantum-mechanical Hamiltonian operator of the system, and the system is here described solely through the qubit Hamiltonian~\cite{deffner_quantum_2019}. Thus, the internal energy of the system reads
\begin{equation}\label{eq:internal_energy}
    E = \langle\hat{H}\rangle = \mathrm{Tr}\left\{\hat{\rho}\hat{H}\right\} =  \sum_n p_n E_n,
\end{equation}
where $E_n$ are the eigenvalues of the qubit, and $p_n$ are the populations of each eigenstate in the qubit. The state of the qubit during the whole thermodynamic cycle can accurately be described as a Gibbs state, a thermal equilibrium state of a given effective temperature $T_\textrm{eff}$. Here, the populations of the system are given by the Boltzmann distribution as $p_n = \mathrm{exp}\left(-\beta E_n\right)/\left[\sum_n \mathrm{exp}\left(-\beta E_n\right)\right]$, where $\beta=1/(k_\textrm{B}T_\textrm{eff})$ and $k_\textrm{B}$ is the Boltzmann constant. We can approximate the effective temperature of the system by fitting the measured populations to this distribution (Methods). 

The first law of thermodynamics states that the increase in the internal energy of the system $\mathrm{d}E$ is given by the sum of the work done on the system $\delta W$ and the heat supplied to the system $\delta Q$~\cite{deffner_quantum_2019}. Thus, by differentiating the internal energy in Eq.~\eqref{eq:internal_energy}, we obtain the quantum version of the first law of thermodynamics for the Otto cycle~\cite{deffner_quantum_2019}
\begin{equation}\label{eq:work_heat}
    \mathrm{d}E = \sum_n\left( p_n \mathrm{d}E_n + E_n \mathrm{d}p_n\right) = \delta W + \delta Q.
\end{equation}

\begin{figure}
  \centering 
  \ifincludefigures\includegraphics[width=0.87\linewidth]{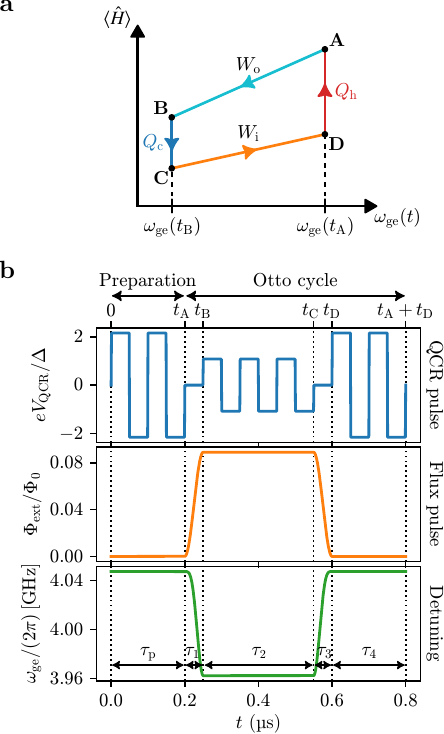}\fi
  \caption{\label{fig:cycle_theory}\textbf{Implementation of the quantum Otto cycle.}~(\textbf{a})~Ideal quantum Otto cycle illustrated in the transition-frequency--mean-energy plane. The thermodynamic processes of the heat engine strokes are displayed with colored lines and the states at the stroke endpoints are denoted with letters A to D. The isochoric heat strokes are realized with QCR AC pulses and the adiabatic work strokes with trapezoidal flux pulses on the sample displayed in Fig.~\ref{fig:setup}a using the pulse shapes depicted in (\textbf{b}). Dashed lines indicate the initial and the detuned lowest-transition frequencies of the qubit. (\textbf{b})~The QCR junction voltage, the normalized external flux, and the lowest-transition frequency of the qubit as functions of time for the preparation stage and the quantum Otto cycle. The junction voltage and external flux depict the pulse shapes for the quantum Otto cycle. The dotted vertical lines with times $t_i$ correspond to the endpoints of the strokes $i$ depicted in (\textbf{a}), and the stroke lengths $\tau_i$ are depicted with two-way arrows.}
\end{figure}

Following the labeling convention defined in Fig.~\ref{fig:cycle_theory}a, we can describe the quantum Otto cycle as follows: The adiabatic expansion (A$\rightarrow$\,B) is determined by the decrease of the transition frequencies and thus eigenenergies of the system, during which the system does work on the field driving the change of the frequencies with no exchange of heat between the system and the environment, implying that $\mathrm{d}p_n = 0$. The isochoric cooling stroke (B$\rightarrow$\,C) is given by the decrease in the state populations owing to a cold reservoir which absorbs thermal energy from the system, thus decreasing the state populations in the excited states. The adiabatic compression (C$\rightarrow$\,D) is given by the increase of the transition frequencies during which the driving field does work on the system. Owing to a smaller number of excitations in the qubit than during the adiabatic expansion, the work done on the system is here smaller than the work done by the system during the expansion. As a final stroke, the isochoric heating (D$\rightarrow$\,A) stroke is given by the increase in the state populations as coupling to a hot bath drives the quantum system to reach a state with a higher effective temperature. During the isochoric strokes, the eigenenergies of the system stay constant with $\mathrm{d}E_n = 0$. This quantum Otto cycle is illustrated in Fig.~\ref{fig:cycle_theory}a in the context of the qubit system, where the energy diagram indicates the change in the eigenfrequency and mean energy during the strokes of the cycle.  

Provided that the populations of the qubit are measured, we can use Eq.~\eqref{eq:work_heat} to obtain the heat absorbed from the effective hot reservoir by the qubit $Q_\mathrm{abs} = Q_\mathrm{h} = \int_{\mathrm{D}\rightarrow\mathrm{A}}\delta Q$ and the total work done by the qubit during the adiabatic strokes $W_\mathrm{tot} = W_\mathrm{i} + W_\mathrm{o}=\int_{\mathrm{C}\rightarrow\mathrm{D}}\delta W + \int_{\mathrm{A}\rightarrow\mathrm{B}}\delta W$. 
Thus, we may characterize the operation of this device by extracting the output power and efficiency as 
\begin{equation}\label{eq:power_efficiency}
    P = \frac{-W_\mathrm{tot}}{\tau_\mathrm{cyc}}, \qquad \eta = \frac{-W_\mathrm{tot}}{Q_\mathrm{abs}},
\end{equation}
where $\tau_\mathrm{cyc}$ is the period of the cycle. In our convention, the negative sign in front of $W_\textrm{tot}$ is used to denote that work is extracted from the qubit for $W_\textrm{tot}<0$, and the positive sign in front of $Q_\textrm{abs}$ implies that heat is absorbed by the qubit for $Q_\textrm{abs}>0$.

\begin{figure*}
  \begin{center}
    \ifincludefigures\includegraphics[width=1\linewidth
    ]{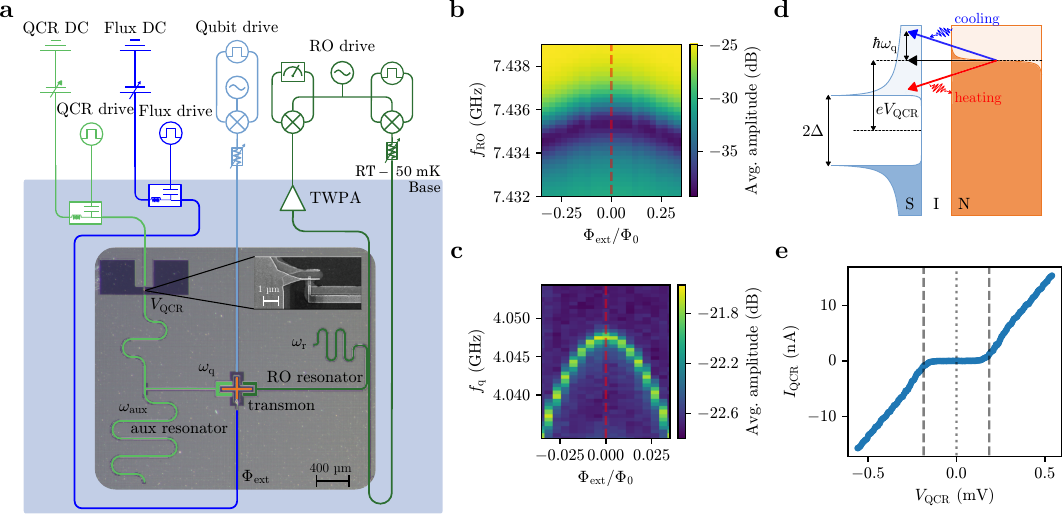}\fi     
  \end{center}
  \caption{\small\label{fig:setup}\textbf{Experimental setup, sample, and their characterization.}~(\textbf{a})~Microscope image of the sample and the circuit diagram of the microwave drive and readout lines. The sample consists of a transmon qubit (orange), a quantum-circuit refrigerator (QCR) coupled to an auxiliary resonator (light green), a qubit drive line (light blue), a qubit flux line (dark blue), and a readout resonator (dark green) on a silicon chip. The inset displays the normal-metal--insulator--superconductor (NIS) junction of the QCR. The circuitry consists of direct-current (DC) and alternating-current (AC) drive lines for the QCR combined with a bias tee, flux DC and AC drive lines, qubit driving circuitry for state preparation, and circuitry for the single-shot readout, along with a traveling-wave parametric amplifier (TWPA). (\textbf{b}), (\textbf{c})~Average amplitude of the microwave tone through the readout resonator as a function of the normalized external flux and (\textbf{b}) readout tone frequency and (\textbf{c}) qubit drive frequency. The flux sweet spots are marked with red vertical dashed lines. The dip in the readout tone and the peak in the qubit drive frequency yield the resonance frequency of the resonator and the lowest-transition frequency of the qubit, respectively. (\textbf{d})~Energy diagram for the photon-assisted tunneling events at the NIS junction of the QCR. The quasiparticle excitations follow the Fermi distribution adopted in the constant density of states of the normal metal (orange) and in the Bardeen--Cooper--Schrieffer density of states of the superconductor (blue), exhibiting a characteristic energy gap of $2\Delta$. The bias voltage $V_\mathrm{QCR}$ shifts the electromagnetic potential of the normal metal and enables the tunneling of quasiparticles from the normal metal to the superconductor with three processes: tunneling by photon absorption (blue arrows), tunneling by photon emission (red arrows), and elastic tunneling (black arrow). The energy of the absorbed and emitted photons is determined by the energy of the corresponding qubit transition $\hbar\omega_\mathrm{q}$. (\textbf{e})~Measured current through the QCR as a function of its bias voltage $V_\mathrm{QCR}$. The dotted vertical line depicts the zero-voltage bias point, to which the curve is normalized, and the dashed vertical lines depict the edges of the subgap region, which provide the $2\Delta$ width of the region.}
\end{figure*}

\subsection*{Experimental setup and sample}
In our experiment, we study a pioneering realization of a quantum heat engine using a sample consisting of a transmon qubit~\cite{koch_charge-insensitive_2007} as the working medium of the thermodynamic cycle coupled to a single-junction QCR and to a capacitively coupled readout resonator for qubit state readout. The microwave pulsing circuitry, the readout circuitry, and the qubit chip, along with its components, are illustrated in Fig.~\ref{fig:setup}a, and a more detailed wiring scheme is displayed in Fig.~\ref{fig:ext_measurement_wiring}. The lowest transition of the qubit is at angular frequency $\omega_\mathrm{ge} = 2\pi\times4.047\,\mathrm{GHz}$ with an anharmonicity of $\alpha = -2\pi\times279\,\mathrm{MHz}$. The qubit is capacitively coupled to the QCR via an auxiliary coplanar-waveguide (CPW) resonator with resonance angular frequency $\omega_\mathrm{aux} = 2\pi\times4.670\,\mathrm{GHz}$. The resonator used for dispersive readout of the qubit state~\cite{blais_cavity_2004,goppl_coplanar_2008} is a CPW resonator with angular frequency $\omega_\mathrm{r} = 2\pi\times7.436\,\mathrm{GHz}$. The resonance frequencies of the superconducting-circuit elements are measured using single- and two-tone spectroscopies displayed in Figs.~\ref{fig:setup}b and \ref{fig:setup}c and explained in more detail in the Methods. 

We describe the qubit using the Hamiltonian~\cite{krantz_quantum_2019}
\begin{equation}\label{eq:qubit_hamiltonian}
    \hat{H}(t) = \hbar\omega_\mathrm{ge}(t)\hat{b}^\dagger\hat{b} + \frac{\hbar\alpha}{2}\hat{b}^{\dagger2}\hat{b}^2,
\end{equation}
where $\hat{b}$ is the annihilation operator of the qubit and $\hat{b}^\dagger$ is the corresponding creation operator. As detailed below, the operation of the heat engine calls for a temporal dependence of the qubit frequency, induced here by the external flux, and of the incoherent transition rates between qubit levels, which we control by the QCR.

The single-junction quantum-circuit refrigerator consists of a normal-metal--insulator--superconductor (NIS) junction galvanically coupled with a direct ohmic connection to an auxiliary reset resonator for circuit coupling~\cite{vadimov_single-junction_2022,viitanen_quantum-circuit_2024}. As indicated by the energy diagram in Fig.~\ref{fig:setup}d, we can induce tunneling of quasiparticles by changing the electrochemical potential of the normal-metal with a bias voltage $V_\mathrm{QCR}$. Voltages in the range $\Delta - \hbar\omega_\mathrm{q} \lesssim e|V_\mathrm{QCR}| \lesssim \Delta$ primarily allow tunneling of quasiparticles from the normal metal to the superconductor through absorption of a photon. Voltages  $e|V_\mathrm{QCR}| > \Delta + \hbar\omega_\mathrm{q}$ also promote elastic tunneling and tunneling through the emission of photons. Importantly, the QCR appears to the qubit accurately as a coupled thermal reservoir, the temperature and coupling strength of which are tunable by the bias voltage~\cite{silveri_theory_2017}. Essentially, the QCR-induced transition rates correspond to a thermal reservoir with a temperature determined by the quasiparticle photon-assisted-tunneling rates as a function of the bias voltage (Methods)~\cite{silveri_theory_2017}. Therefore, we may employ these two regimes to effectively cool or heat the transmon to thermal states with lower or higher effective temperatures, respectively~\cite{teixeira_many-excitation_2024,morstedt_rapid_2025}. 

The QCR is characterized by measuring the current-voltage (IV) curve of the device displayed in Fig.~\ref{fig:setup}e, from which we can infer the superconducting energy gap parameter $\Delta = 186\,\text{\textmu eV}$ as the half-width of the zero-voltage gap region, the tunneling resistance $R_\mathrm{T} = 25.7\,\mathrm{k\Omega}$ from the differential resistance at high $V_\mathrm{QCR}$, and the Dynes parameter $\gamma_\mathrm{D} = 4.0\times10^{-3}$ depicting the density of subgap states~\cite{silveri_theory_2017,hsu_tunable_2020} as the ratio of the tunneling resistance and the subgap resistance. The values of the measured characteristic parameters of the sample are compiled in Table~\ref{tab:parameters}. 

\begin{table}\footnotesize
  \caption{\label{tab:parameters}\textbf{Summary of the measured device parameters.} The experimental measurement protocols for determining the parameters are explained in detail in the Methods.}
  \begin{tabular}{l|c|c}
    \hline
    Parameter & Symbol & Value\\  
    \hline    
    \hline
    Readout resonator angular frequency &  $\omega_{\rm r}$& $2\pi\times$\qty{7.436}{\GHz} \\
    Reset resonator angular frequency &  $\omega_{\rm aux}$ & $2\pi\times$\qty{4.670}{\GHz} \\
    Qubit angular frequency & $\omega_{\rm ge}$ & $2\pi\times$\qty{4.047}{\GHz} \\
    Qubit anharmonicity  &  $\alpha$ & $-2\pi\times$\qty{279}{\MHz}\\ 
    Tunneling resistance  & $R_{\rm T}$ & \qty{25.7}{\kohm}\\
    Superconductor gap  & $\Delta $ & \qty{186}{\micro\eV} \\
    Dynes parameter  &$\gamma_{\rm D}$ & \num{4.0e-3} \\
    \hline 
  \end{tabular}
\end{table}

\subsection*{Experimental protocol and calibration}
Experimentally, the adiabatic strokes are implemented by tuning the eigenenergies of the qubit by varying the external flux through the SQUID loop of the transmon. This is achieved by sending trapezoidal current pulses through the flux line of the sample, shown in Fig.~\ref{fig:cycle_theory}b. The ramp-up and down sections of the pulses adiabatically tune the eigenfrequencies of the qubit as defined by the flux-tunable transmon eigenenergies $E_m = \hbar\omega_\mathrm{ge}m + \hbar\alpha\left(m^2 - m\right)/2$. The lowest-transition angular frequency is given by $\omega_\mathrm{ge} = \left(\sqrt{8E_\mathrm{C}E_\mathrm{J}(\Phi_\mathrm{ext})} - E_\mathrm{C}\right)/\hbar$, where $E_\mathrm{C}$ is the charging energy of the transmon, and $E_\mathrm{J}(\Phi_\mathrm{ext}) = E_\mathrm{J}^\mathrm{max}|\cos(\Phi_\mathrm{ext}/\Phi_0)|$ is the flux tunable Josephson energy of the transmon~\cite{krantz_quantum_2019}. Essentially, the flux tuning changes the inductive energy of the qubit circuit, and therefore, the magnetic field generated by the flux line does work on the system, and the system does work on the magnetic field during the adiabatic compression and expansion strokes, respectively. Since in the transmon regime, the qubit Hamiltonian is dominated by the Josephson Hamiltonian, which is proportional to  $E_\textrm{J}(\Phi_\textrm{ext})\gg 1$~GHz, the adiabaticity of the flux sweeps is well satisfied for the sweep times of tens of nanoseconds employed in this work.

\begin{figure*}
  \centering 
  \ifincludefigures\includegraphics[width=0.94\linewidth
  ]{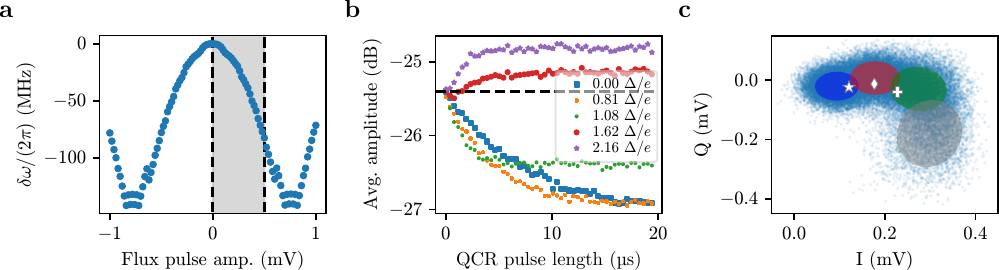}\fi
  \caption{\label{fig:cycle_calibration}\textbf{Calibration of qubit frequency, QCR bias, and single-shot readout for the Otto cycles.}~(\textbf{a})~Frequency detuning of the lowest-transition frequency of the qubit obtained from a modified Ramsey measurement as a function of the pulse amplitude of a square flux pulse. The dashed lines and the grey shading indicate the used detuning range for the heat engine measurements with a maximum $\delta\omega = -2\pi\times82.4\,\mathrm{MHz}$ detuning. (\textbf{b})~Trajectory average amplitude of the microwave tone through the readout resonator as a function of QCR pulse length for multiple pulse amplitude values. The qubit state is prepared in the excited state before measurement, and the horizontal dashed line depicts the initial amplitude at the excited state. The applied QCR pulse is a net-zero square pulse with a period of 100\,ns. (\textbf{c})~Single-shot calibration data from 10\,000 shots of prepared $\ket{\rm g}$, $\ket{\rm e}$, and $\ket{\rm f}$ states of the qubit each, together with fitted two-dimensional Gaussian distributions of elliptical shapes for ~$\ket{\rm g}$ (blue), $\ket{\rm e}$ (red), and $\ket{\rm f}$ (green). In addition, the grey distribution corresponds to high-lying states outside the other distributions denoted as $\ket{\rm h,i,j}$. The points marked with a star, a diamond, and a cross depict the mean values of the $\ket{\rm g}$, $\ket{\rm e}$, and $\ket{\rm f}$ calibration data sets, respectively, highlighting the difference between the means of the data and the Gaussian fits.}
\end{figure*}

For the isochoric strokes, we implement the heating and cooling of the transmon circuit by biasing the QCR junction with pulses of alternating current (AC) of varying amplitude to allow for the different tunneling effects to be emphasized at will and to generate thermal states in the transmon qubit with desired effective temperatures determined by the amplitude~\cite{teixeira_many-excitation_2024,morstedt_rapid_2025}. The shapes of the pulses are depicted in Fig.~\ref{fig:cycle_theory}b, where the QCR bias voltage and SQUID flux are displayed as a function of time. The QCR pulse is a net-zero square pulse with a period of 100\,ns, whereas the flux pulse is a trapezoidal pulse with a ramp-up time matching the length of the adiabatic stroke. The QCR pulse and the flux pulse are sent simultaneously, and always one of the pulses remains unchanged for the stroke duration of the other pulse. The figure also displays how the flux pulse tunes the transition frequency, matching the frequency evolution in the Otto cycle shown in Fig.~\ref{fig:cycle_theory}a.

To prepare for the heat engine measurements, we characterize the behavior of the qubit state under the AC pulses and calibrate the single-shot measurement technique with a Gaussian mixture model (GMM) used for qubit state identification from single shots (Methods). The flux pulse amplitude is characterized so that the detuning of the qubit is known for a given amplitude during the cycle. This is carried out by a modified Ramsey measurement~\cite{krantz_quantum_2019} where we set the detuning of the qubit drive to zero and modulate the qubit detuning by sending a square flux pulse during the idle time of the Ramsey measurement. Due to the flux pulse, the idle time is fixed for the Ramsey measurement. Instead, the relative phase of the second $\pi/2$ pulse of the measurement is swept to gain Ramsey fringes. By sweeping the flux pulse amplitude, we obtain the qubit detuning $\delta\omega$ from the fit parameters of the Ramsey fringes as a function of flux pulse amplitude, displayed in Fig.~\ref{fig:cycle_calibration}a. Combining the obtained detuning values and the measured qubit transition frequencies, we gain the eigenfrequencies of the qubit along the adiabatic strokes of the Otto cycle.

The strengths of the cooling and heating strokes are characterized before the Otto cycle measurements by exciting the qubit to the excited state with a $\pi$ pulse, sending a net-zero square pulse through the QCR junction, and varying the length of this pulse for multiple amplitudes. This calibration determines the average amplitude of the readout signal, to which the state reaches for the given amplitude of the QCR drive, which can be used as a rough indicator for the effective temperature of the qubit without measuring the distribution using single shots. The calibration data shown in Fig.~\ref{fig:cycle_calibration}b demonstrate that the low pulse amplitudes of $0.81\,\Delta/e$ and $1.08\Delta/e$ decrease the average amplitude of the readout tone faster than the bare decay with $0.0\,\Delta/e$ amplitude. Although the amplitude $1.08\,\Delta/e$ is slightly over the superconducting gap parameter and leads to higher qubit excitation than the bare decay, it yields the fastest decrease at short time scales comparable to those used in the main measurements described below. In this regime, the QCR-induced qubit relaxation rate is increased while the excitation rate stays low enough, leading to faster initial decay compared to that for lower amplitudes~\cite{teixeira_many-excitation_2024,hsu_tunable_2020}. For amplitudes greatly exceeding the superconducting gap parameter, the average amplitude of the readout tone increases asymptotically with a higher pulse amplitude corresponding to a higher saturated readout amplitude. 

\begin{figure*}
  \centering 
  \ifincludefigures\includegraphics{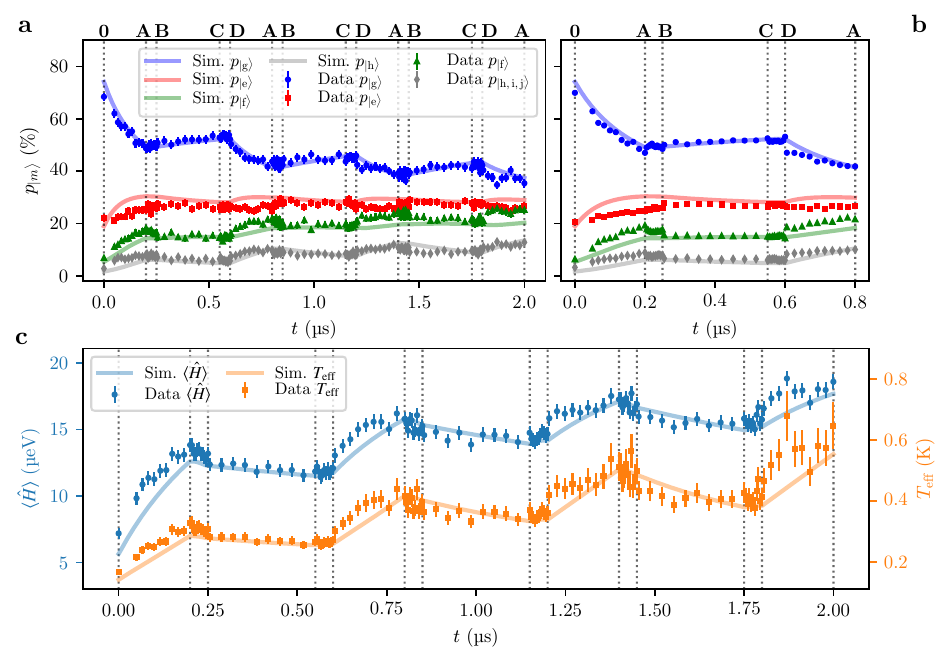}\fi
  \caption{\label{fig:main_results}\textbf{Quantum heat engine dynamics.}~(\textbf{a})~Measured state populations (markers) as functions of time for three consecutive Otto cycles starting from the thermal state obtained without any pulses applied. The measured populations are obtained from single-shot distributions, whereas the solid lines depict corresponding simulated values using the system parameters of Table~\ref{tab:parameters}. The dotted vertical lines display the endpoints of each stroke displayed at the top of the panel. The $1\sigma$ error bars are obtained from the standard deviation of the population data. (\textbf{b})~Experimentally obtained (markers) and simulated (solid lines) mean state populations as functions of time for eight repetitions of the first Otto cycle starting from the thermal ground state. The $1\sigma$ error bars, obtained from the combined standard error of the repetitions, are smaller than the marker size.  (\textbf{c})~Mean energy of the transmon (left vertical axis) and the temperature corresponding to the nearest thermal state (right vertical axis) as functions of time for three consecutive Otto cycles obtained from the data of panel (\textbf{a}) and the energies of each transmon state during the cycle (Methods). The $1\sigma$ error bars are obtained from the standard deviation of the mean energy via standard error propagation, and the error of the temperature is acquired using Monte Carlo sampling from the measured populations and taking the standard deviation of the temperatures from fitting these samples (Methods).}
\end{figure*}

The qubit state measurements are carried out with single shots, and the state identification is realized using a multivariate Gaussian mixture model (Methods). To calibrate this, we prepare the qubit state to $\ket{\rm g}$, $\ket{\rm e}$, and $\ket{\rm f}$ using corresponding $\pi$ pulses, and measure 10\,000 shots from each state to a combined calibration dataset. The combined single shots representing the point clouds of the three lowest-energy states of the qubit are fitted with four two-dimensional Gaussian distributions, with the fourth distribution accounting for sporadic higher excitations in the qubit state. The calibration data set with the fitted GMM distributions is displayed in Fig.~\ref{fig:cycle_calibration}c. We observe a larger overlap of the data points and the distributions than in previous measurements with similar samples in Refs.~\cite{teixeira_many-excitation_2024,morstedt_rapid_2025} due to lower signal-to-noise ratio and coherence times. However, the point clouds remain separated enough for the GMM fitting. The qubit state populations are calculated in the following measurements by first counting the number of single shots inside the $0.4\boldsymbol{\sigma}$-boundary ellipses of each state distribution and then, to account for the overlap of the point clouds, we employ a correction matrix to the counted single shots before normalizing the counts with the total number of corrected shots (Methods).

\subsection*{Experimental demonstration}
The pulse lengths for measuring the heat engine dynamics are $\tau_\mathrm{p} = \tau_4 = 200\,\mathrm{ns}$, $\tau_1=\tau_3 = 50\,\mathrm{ns}$, and $\tau_2 = 300\,\mathrm{ns}$ as displayed in Fig.~\ref{fig:cycle_theory}b. Here, the cooling pulse is the limiting factor for the cycle time,
since the related decay and excitation rates are low in comparison to the adiabaticity condition for the flux sweeps. The QCR pulse amplitudes are $A_\mathrm{h} = 2.16\,\Delta/e$ for the heating pulse amplitude and $A_\mathrm{c} = 1.08\,\Delta/e$ for the cooling pulse amplitude. The flux amplitude is $A_\mathrm{f} = 0.5\,\mathrm{mV}$ corresponding to a detuning of $\delta\omega = -2\pi\times82.4\,\mathrm{MHz}$ which is selected from the highest detuning values from Fig.~\ref{fig:cycle_calibration}a without reaching the half-flux region where the cosine term of the Josephson energy approaches zero and where the flux calibration has a large variance. With these cycle parameters and starting from the thermal qubit state without biasing the QCR, we measure the transmon state populations for three consecutive cycles realized with the above-described flux and QCR pulses. We choose to closely study only three cycles for these high-resolution sweeps owing to limitations in the measurement time, driving pulse lengths, and the required database size of each run. Each data point along the cycle consists of populations inferred from 10\,000 time-integrated single shots using a 2\,\textmu s readout pulse. The measured state populations are displayed in Fig.~\ref{fig:main_results}a for the four states $|\mathrm{g\rangle}, |\mathrm{e\rangle}, |\mathrm{f\rangle},\ \text{and}\ |\mathrm{h,i,j\rangle}$ used in the GMM model starting from the thermal state with effective temperature $T_\mathrm{int} = 160$\,mK. In addition, we simulate the dynamics of the transmon populations using a Lindblad master equation (Methods) with the corresponding system, pulse, and cycle parameters. The measured populations agree with the simulated values, and we observe cyclic changes in the measured populations as expected. The changes are larger in the first cycle, and we start reaching a relatively well-saturated thermal-state distribution at the end of the third cycle.

From the measured transmon populations, we compute the mean energy of the qubit using the measured transition frequencies for the strokes of the cycle and Eq.~\eqref{eq:internal_energy}. We also estimate the effective temperature of the transmon by fitting the measured populations to the Boltzmann distribution describing the populations of a Gibbs state and by finding the effective qubit temperature from the fit parameters (Methods). These are displayed in Fig.~\ref{fig:main_results}c together with values calculated from the populations in the simulated dynamics. We observe that the experimentally obtained mean energy and effective temperature agree well with the simulations. We observe an increase in the effective temperature from around 200\,mK to 600\,mK in three cycles, indicating that the heating stroke is not fully compensated by the cooling stroke at low effective temperatures, resulting in a temperature increase for low initial effective temperatures.

Even though these experimental results agree well with the simulated values and the demonstrated cycle accurately implements the quantum Otto cycle, the cycle is not perfect since the mean energies at the beginning and end of the cycle do not exactly match for any of the cycles. Fortunately, we observe that this mismatch decreases with increasing effective temperature, and eventually the state will saturate and approach a steady-state operation regime. This behavior is demonstrated in our simulations in Fig.~\ref{fig:ext_saturation_simulations}, where the cycle approaches an ideal Otto cycle in this sense over time. This steady-state operation may be reached by decreasing the amplitude and length of the heating stroke, by increasing those of the cooling stroke, by measuring later cycles, or by starting the cycle with a strong heating pulse that initializes the effective temperature to the steady-state regime. Nevertheless, we realize a quantum heat engine cycle even in Fig.~\ref{fig:main_results}c since the work done to the system is less than the work done by the system to the flux environment, thus leading to positive output power. Note also that the effective temperature of the qubit at the end of the third cycle is almost equal to that after the second cycle. Although quantum coherences are not expected to have a direct influence on the operation of the realized quantum heat engine, which functions primarily on incoherent mixed states, this superconducting circuit framework can be combined with quantum state tomography~\cite{liu_performing_2023} to serve as an experimental platform to investigate the operation of quantum heat engines on coherently superposed states.

To further characterize the implemented Otto cycle, we carry out repeated measurements of the first Otto cycle starting from the thermal state without pulses. We measure and compute the characteristic thermodynamic quantities of the cycles and compare them to those acquired from simulations. The mean values of the measured state populations and their uncertainties from eight repetitions are shown in Fig.~\ref{fig:main_results}b along with the simulated dynamics of the populations. Again, we observe an agreement with the simulations and a small variance between the measurements. Using the definitions of work and heat from Eq.~\eqref{eq:work_heat}, we numerically integrate the strokes using the measured populations, transition frequencies, and detunings from Figs.~\ref{fig:main_results}a, \ref{fig:setup}c, and \ref{fig:cycle_calibration}a to compute the work terms $W_\mathrm{o}$ and $W_\mathrm{i}$, and heat terms $Q_\mathrm{h}$ and $Q_\mathrm{c}$. Consequently, we obtain the output power $P = -W_\mathrm{tot}/\tau_\mathrm{cyc}$ and the efficiency $\eta = -W_\mathrm{tot}/Q_\mathrm{abs}$ for each repetition of the cycle. The mean values of these quantities and of the total work and absorbed heat are shown in Table~\ref{tab:therm_values} together with their experimental uncertainties and corresponding simulated values. The relative uncertainties lie below 5\%, indicating that the fluctuations of the quantities during the cycle are moderate compared to their mean values, consistent with the quasi-static nature of the strokes.

From the perspective of stochastic thermodynamics, miniaturized heat engines are expected to exhibit fluctuations in their output, which are constrained by the thermodynamic uncertainty relations (TURs)~\cite{landi_irreversible_2021}. In our experiment, we access the statistics of the engine output by repeating several Otto-cycle measurements and extracting thermodynamic quantities for each repetition, finding relatively small variances compared to the mean values. A quantitative TUR test would require a trajectory-level description of stochastic heat currents and entropy production, which lies beyond the scope of the present work. We therefore restrict ourselves to characterizing ensemble-level fluctuations of the measured quantities.

\begin{table}
  \caption{\label{tab:therm_values}\textbf{Summary of the measured performance of the quantum heat engine.} The mean, standard error of the mean, and simulated values of the absorbed heat $Q_\mathrm{abs}$, total work $W_\mathrm{tot}$, output power $P$, and efficiency $\eta$ from eight repetitions of the first Otto cycle. The experimental measurement protocols for determining the parameters are detailed in the Methods.}
  \vspace{1mm}
    \begin{tabular}{c|c|c|c}
            \hline
            Parameter   & Mean   & Standard error     &   Simulated \\ 
            \hline
            \hline
            $Q_\mathrm{abs}$  & \qty{4.22}{\micro\eV}   & \qty{0.14}{\micro\eV}     & \qty{4.06}{\micro\eV}       \\
            $W_\mathrm{tot}$      & \qty{-0.023}{\micro\eV}  & \qty{0.0010}{\micro\eV}   & \qty{-0.018}{\micro\eV}       \\
            $P$       & \qty{0.039}{\eV/s}  & \qty{0.0017}{\eV/s} & \qty{0.031}{\eV/s}       \\
            $\eta$   & 0.0055    &  0.0004    & 0.0045      \\\hline
    \end{tabular}
\end{table}

The power and efficiency are mostly limited by the finite detuning range shown in Fig.~\ref{fig:cycle_calibration}a since the efficiency is expected to be in the same range as for a quantum Otto engine with an ideal quantum harmonic oscillator which has the efficiency $\eta_\mathrm{Otto} = 1 - \kappa = 1 - \omega_\mathrm{min}/\omega_\mathrm{max}$ where $\kappa$ is referred to as the compression ratio~\cite{deffner_quantum_2019,kosloff_quantum_2017}. Thus the maximum detuning of $-82.4\,\mathrm{MHz}$ used in the experiments yields an Otto efficiency of $\eta_\mathrm{Otto} = 1 - \omega_\mathrm{ge}(t_\mathrm{B})/\omega_\mathrm{ge}(t_\mathrm{A}) \approx 0.020$~\cite{morstedt_rapid_2025}. The mean output power of the repeated cycles is 0.039\,eV/s, and the mean efficiency is 0.0055, which is 27~\% of the ideal Otto efficiency, reproduced relatively accurately by the simulated values 0.031\,eV/s and 0.0045, respectively. Since the first cycle is far from saturation, the corresponding absorbed heat is significantly higher than for the steady-state operation, which accounts for the reduced efficiency of $\eta = 0.27 \eta_\mathrm{Otto}$. Thus, the efficiency is lower for the first cycle, but the value quickly approaches the Otto efficiency during the first few cycles. Although the acquired efficiency and power are somewhat low and currently lack practical applications, they provide very strong evidence for our main result that the cycle implemented by the AC pulses demonstrates a quantum heat engine with a positive output power and efficiency, thus generating work from the heat flowing through the transmon. Furthermore, the cycle efficiency of the system is predicted by our simulations to reach the Otto efficiency regime in the steady-state operation, as discussed in the Methods. In the existing literature~\cite{peterson_experimental_2019,bouton_quantum_2021}, efficiencies of 0.42--0.48 have been reported for non-superconducting QHEs, but these values are also governed by Otto efficiency, although with high compression ratios.

\section*{Discussion}
In conclusion, we have experimentally demonstrated a quantum Otto heat engine with superconducting circuits by operating a single-junction quantum-circuit refrigerator as a temperature-tunable reservoir for a flux-tunable transmon qubit. For the first few cycles, we achieved positive output powers and efficiencies exceeding 25\% of the Otto efficiency. Our results agree with those obtained from numerical simulations of an open quantum system similar to the experimental sample, driven by identical AC pulses. This work provides an initial proof of concept for controlling the thermal environment and magnetic flux of a superconducting qubit, enabling the realization of a quantum heat engine using superconducting circuits. Curiously, we utilize a non-conventional single source for the thermal reservoirs of the heat engine that can be turned on and off at will, and demonstrate the capabilities of accurately measuring the quantum state of the engine at different points of the cycle. 

Our results provide a verification scenario for theoretical treatments of quantum thermodynamics and especially of quantum heat engines, possibly paving the way for future optimized heat engine realizations and other applications of thermally controlled superconducting circuits. In the future, the dynamics of this device may be further studied by including the reset resonator dynamics in the theoretical model and by operating the engine cycle while applying additional drives to the reset resonator and/or the qubit to study the effects of quantum coherences or even exceptional points, as in Ref.~\cite{zhang_dynamical_2022}. Furthermore, the differences between cyclic and non-cyclic heat engine regimes may be investigated, together with non-adiabatic work strokes to study the effect of quantum interference and related Landau-Zener-Stückelberg transitions.

\section*{Methods}

\subsection*{Sample fabrication}
The sample in Fig.~\ref{fig:setup}a is fabricated using a series of micro- and nanofabrication techniques. The ground plane, the resonators, and the cross shape of the transmon are fabricated on a six-inch intrinsic-silicon wafer using optical lithography and reactive ion etching of a 200-nm film of sputtered niobium. The Josephson junctions in the superconducting quantum interference device (SQUID) of the transmon are added using electron beam lithography (EBL) by tailoring an undercut in a double-layer resist of a methyl methacrylate copolymer and polymethyl methacrylate, and then depositing two 30-nm aluminum layers using a two-angle shadow evaporation technique with in-situ oxidation for the insulating barrier in the SIS junction. The NIS junction of the QCR is fabricated similarly by replacing the top aluminum layer of the SIS junction with 60~nm of copper, which has an additional 3-nm adhesion layer of Al to reduce oxidation and diffusion of the copper without introducing superconductivity.

\begin{figure*}
  \centering 
  \ifincludefigures\includegraphics[width=\linewidth]{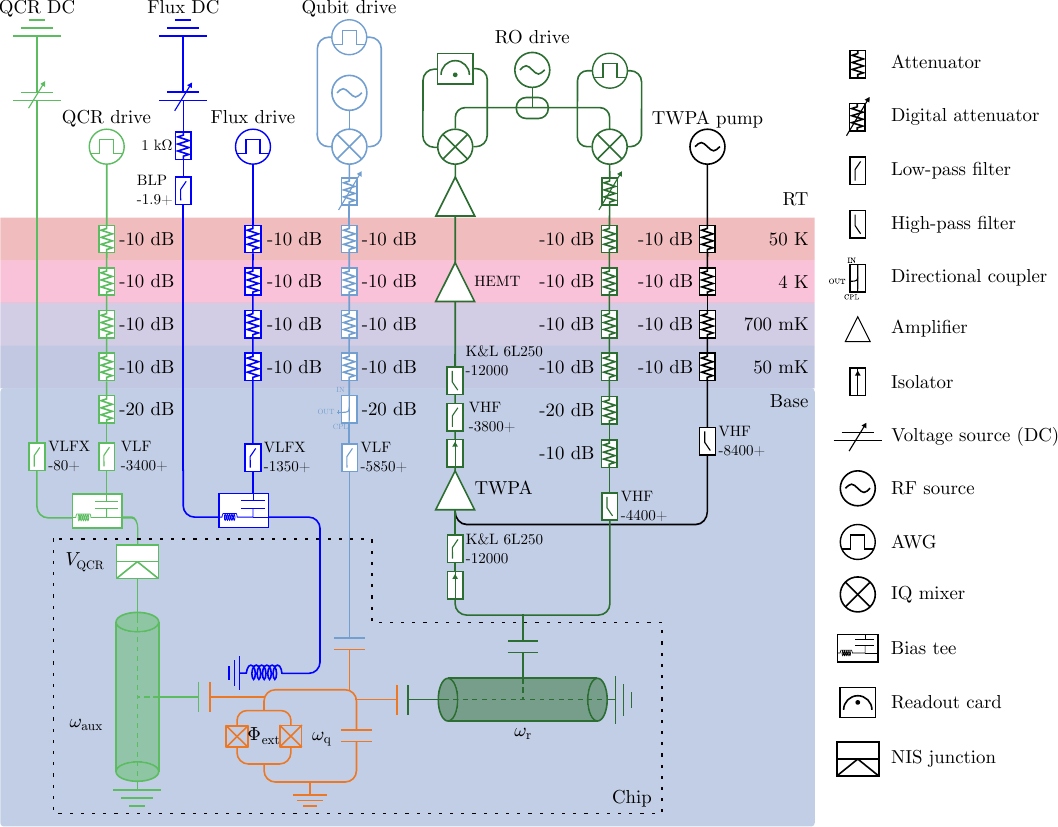}\fi
  \caption{\label{fig:ext_measurement_wiring}\textbf{Circuit diagram of the measurement setup wiring.} Detailed circuit diagram of the measurement setup displaying the microwave drive lines and readout lines with their corresponding attenuators and filters. The circuit of the sample chip is illustrated using the same coloring scheme as in Fig.~\ref{fig:setup}a.}
\end{figure*}

\subsection*{Measurements}
For all of the measurements, we use a commercial dilution refrigerator to cool down the sample to a base temperature of around 40~mK. The qubit chip is adhered to a copper sample holder, and the microwave lines on the chip are bonded to the transmission lines on the holder with aluminum bonding wires. Airbridges are created over the waveguides of the chip using bonding wires between ground plane sections to reduce unwanted resonances, such as slotline modes. The holder is packaged into a magnetic shield to reduce interference from stray magnetic fields on the SQUID loop of the transmon. The transmission line ports in the sample are connected to room-temperature microwave equipment with coaxial cables, attenuators, and amplifiers as displayed in the measurement wiring scheme in Fig.~\ref{fig:ext_measurement_wiring}.

The direct-current (DC) bias voltage on the NIS junction of the QCR is applied using a source measure unit, which simultaneously measures the current through the junction to gain the IV characteristics of the junction shown in Fig.~\ref{fig:setup}e. The DC bias voltage for the flux tuning of the qubit is implemented using a battery-powered voltage source connected to a resistor and a magnetic coil attached to the sample holder. The DC bias of the QCR is set to the center point of the IV curve, and the DC bias of the flux is set to the flux sweet spot for the QHE measurements. The AC pulses for both the QCR and the flux lines are generated using arbitrary-wave generators (AWGs), which are combined with the DC signals using bias tees.

The trajectory average amplitude and single-shot measurements are carried out using the dispersive shift of the readout resonator frequency by measuring the quadratures of the microwave tone sent through a transmission line coupled to the readout resonator. The 2\,\textmu s readout signal tone is generated by an AWG and digitized by a readout card after separating the quadratures with an IQ mixer. The analog signal is converted to a digital signal from which the qubit-state-dependent in-phase (I) and quadrature-phase (Q) components of the signal are time-integrated. This procedure is repeated 10\,000 times for separate measurements to generate single-shot point clouds in the IQ plane. For trajectory average amplitude measurements, we further average by calculating the average signal amplitude and phase from the 10\,000 time-integrated points.

\subsection*{Device and pulse parameters}
The characteristic parameters of the device are extracted from single- and two-tone spectroscopies and from the IV curve of the QCR junction. The resonance frequency of the readout resonator is obtained by measuring the average amplitude of the microwave tone sent through the readout line for different tone frequencies while tuning the SQUID flux with a DC bias voltage. As shown in Fig.~\ref{fig:setup}c, there is a dip in the measured amplitude when in resonance with the fundamental mode due to the mode being populated by photons from the microwave tone. 

For the QCR characterization, the current through an NIS junction is given by the expression~\cite{karimi_low_2022}
\begin{align}\label{eq:QCR_current}
    I(V,T_\mathrm{N}) = \frac{1}{2 e R_\mathrm{T}}\int_{-\infty}^\infty \mathrm{d}\varepsilon\ n_\mathrm{S}(\varepsilon)\big[&f(\varepsilon-eV, T_\mathrm{N})\nonumber\\
    - &f(\varepsilon + eV, T_\mathrm{N})\big] ,
\end{align}
where the $V$ is the voltage across the junction, $T_\mathrm{N}$ is the temperature of the normal metal island, $R_\mathrm{T}$ is the tunneling resistance, $f(\varepsilon)$ is the Fermi--Dirac distribution, and $n_\mathrm{S}(\varepsilon)$ is normalized density of states given by the expression~\cite{silveri_theory_2017}
\begin{equation}\label{eq:norm_DOS}
    n_s(\varepsilon) = \left|\mathrm{Re}\left\{\frac{\varepsilon + \mathrm{i}\gamma_\mathrm{D}\Delta}      {\sqrt{(\varepsilon + \mathrm{i}\gamma_\mathrm{D}\Delta)^2 - \Delta^2}}\right\}\right|,
\end{equation}
where $\gamma_\mathrm{D}$ is the Dynes parameter and $\Delta$ is the superconductor gap parameter as defined above. The tunneling resistance $R_\mathrm{T}$ is obtained from the slope of the IV curve at the Ohmic region for $eV > \Delta$. The superconducting gap parameter is obtained from the width of the subgap region $2\Delta$, where the current through the junction is nearly zero. This arises from the Bardeen--Cooper--Schrieffer gap where the density of states is effectively zero~\cite{bardeen_theory_1957}. However, owing to, for example, the lifetime broadening of the quasiparticles inside the gap, the density of states is non-zero as given by Eq.~\eqref{eq:norm_DOS}, and it is characterized by the Dynes parameter $\gamma_\mathrm{D}$~\cite{dynes_direct_1978}. This parameter is approximated by the ratio between the tunneling resistance and the resistance at the subgap region with near-zero current. 

The qubit transition frequencies are extracted from two-tone spectroscopies where the average amplitude of the readout tone is measured while applying a continuous drive tone to the qubit drive line with varying frequencies and while tuning the bias voltage of the flux line. The continuous qubit drive will excite the qubit at resonance with the qubit transition frequency, giving a dispersive shift to the readout tone and increasing the readout amplitude. This shift is visible as a peak in the spectroscopy as seen in Fig.~\ref{fig:setup}c. The measurements are carried out for the lowest and higher transition frequencies, while applying the flux voltage sweep, to obtain the transition frequencies at the flux sweet spot corresponding to the initial flux state for the heat engine cycle. These results are used to compute the measured mean energies along the cycle, together with the measured state populations of the qubit.

The QCR pulse amplitudes are characterized for the Otto cycle using a pulse-length sweep measurement. The measurement is carried out by exciting the qubit state to the excited state and measuring the average amplitudes of the readout tone after sending a net-zero square pulse of the selected amplitude for varying pulse lengths. The measurements aim to depict the evolution of the average qubit state during dissipation through the QCR junction for the selected amplitude. The amplitudes are swept for selected ranges to find suitably large and fast decrease and increase of the signal amplitude to implement the cooling and heating pulses, respectively, with a large difference in the induced effective temperature of the qubit. 

For the calibration of the flux-pulse amplitudes, we implement a modified Ramsey measurement where we send a sequence of pulses to the qubit and to the flux line and measure the average amplitude of the readout signal. We start with a $\pi$/2 pulse to excite the qubit to an equal superposition state at the equator of the Bloch sphere. Subsequently, we send a square flux pulse through the flux line that detunes the qubit and let the state evolve for the duration of the flux pulse of length 50\,ns. After this, we send another $\pi$/2 pulse that projects the state either to the ground or to the excited state with a probability depending on the evolution of the qubit during the flux pulse. The probability is dependent on the detuning of the qubit given by the flux pulse amplitude. To obtain the detuning value from this measurement, we sweep the relative phase offset of the second $\pi$/2 pulse to simulate the time evolution of the qubit state in the equator. We measure the average amplitude of the readout pulse sent after the second $\pi$/2 pulse to examine the evolution of the readout amplitude in the Ramsey measurement. The evolution of the qubit state in terms of the averaged readout amplitude $\overline{A}_\mathrm{traj}$ can be approximated by the equation
\begin{equation}\label{eq:mod_Ramsey_fit}
    \overline{A}_\mathrm{traj}(\phi) \propto \cos(\delta\omega\tau + \phi)\mathrm{e}^{-\tau/T_2} + C,
\end{equation}
where $\phi$ is the relative phase offset, $\delta\omega = |\omega_\mathrm{q} - \omega_\mathrm{q,flux}|$ is the detuning of the qubit frequency with the flux pulse, $\tau$ is the length of the flux pulse, $T_2$ is the characteristic transverse relaxation time, and $C$ is a constant offset of the amplitude. We carry out the modified Ramsey measurements along with a sweep of the flux pulse amplitude to obtain the detuning of the qubit from the fit parameters to Eq.~\eqref{eq:mod_Ramsey_fit} for the given pulse amplitude.

\subsection*{Theoretical model}
We model the dynamics of the transmon by a Lindblad master equation employing the following assumptions: We completely ignore the effects of the readout resonator on the transmon dynamics and consider only the effects of the auxiliary resonator through the QCR-induced transition rates of the transmon states~\cite{hsu_tunable_2020}. In general, we assume that the transmon is a finite-dimensional system up to six levels with parameters consistent with previous measurements on the same transmon-QCR system~\cite{teixeira_many-excitation_2024,morstedt_rapid_2025}. We describe the incoherent emission and absorption processes in the master equation as combinations of the intrinsic and QCR-induced baths while omitting the effect of Lamb shifts. In addition, we neglect pure dephasing of the qubit state as the transmon is expected to be in almost fully incoherent mixed states of the energy eigenbasis during the Otto dynamics.

For the frequency-tunable transmon Hamiltonian with time-dependent flux
\begin{equation}
    \hat{H}(t) = \hbar\omega_\mathrm{ge}(t)\hat{b}^\dagger\hat{b} + \frac{\hbar\alpha}{2}\hat{b}^{\dagger2}\hat{b}^2,
\end{equation}
we write the Lindblad master equation for the duration of the preparation and the Otto cycles as~\cite{breuer_theory_2002}
\begin{equation}\label{eq:master_eq}
    \frac{\mathrm{d}\hat{\rho}}{\mathrm{d}t} = -\frac{\mathrm{i}}{\hbar}\left[\hat{H}(t), \hat{\rho}\right] + \mathcal{L}_\downarrow(\hat{\rho}) + \mathcal{L}_\uparrow(\hat{\rho}),
\end{equation}
where $\hat{\rho}$ is the density operator of the system and $\mathcal{L_{\downarrow/\uparrow}}$ are the Liouvillian superoperators describing the dissipative processes of the system. The thermal emission and absorption processes are respectively described by the superoperators
\begin{align}\label{eq:dissip_emission}
    \mathcal{L}_\downarrow(\hat{\rho}) &= \Gamma_\mathrm{eg}(t)\mathcal{D}(\ket{\mathrm{g}}\!\bra{\mathrm{e}})\hat{\rho} + \Gamma_\mathrm{fe}(t)\mathcal{D}(\ket{\mathrm{e}}\!\bra{\mathrm{f}})\hat{\rho}\nonumber\\ 
    &+ \Gamma_\mathrm{hf}(t)\mathcal{D}(\ket{\mathrm{f}}\!\bra{\mathrm{h}})\hat{\rho} + \Gamma_\mathrm{ih}(t)\mathcal{D}(\ket{\mathrm{h}}\!\bra{\mathrm{i}})\hat{\rho}\nonumber\\ 
    &+ \Gamma_\mathrm{ji}(t)\mathcal{D}(\ket{\mathrm{i}}\!\bra{\mathrm{j}})\hat{\rho},
\end{align}
and
\begin{align}\label{eq:dissip_absorption}
    \mathcal{L}_\uparrow(\hat{\rho}) &= \Gamma_\mathrm{ge}(t)\mathcal{D}(\ket{\mathrm{e}}\!\bra{\mathrm{g}})\hat{\rho} + \Gamma_\mathrm{ef}(t)\mathcal{D}(\ket{\mathrm{f}}\!\bra{\mathrm{e}})\hat{\rho}\nonumber\\
    &+ \Gamma_\mathrm{fh}(t)\mathcal{D}(\ket{\mathrm{h}}\!\bra{\mathrm{f}})\hat{\rho} + \Gamma_\mathrm{hi}(t)\mathcal{D}(\ket{\mathrm{i}}\!\bra{\mathrm{h}})\hat{\rho}\nonumber\\
    &+ \Gamma_\mathrm{ij}(t)\mathcal{D}(\ket{\mathrm{j}}\!\bra{\mathrm{i}})\hat{\rho},
\end{align}
where $\Gamma_{nm}(t)$ is the time-dependent transition rate from the transmon state $\ket{n}$ to $\ket{m}$ and the superoperator $\mathcal{D}(\hat{O}) = \hat{O}\hat{\rho}\hat{O}^\dagger - \{\hat{O}^\dagger\hat{O},\hat{\rho}\}/2$ describes the exchange of qubit states~\cite{teixeira_many-excitation_2024}.

The angular eigenfrequencies of the transmon level $\ket{m}$ can be accurately approximated by 
\begin{equation}
    \omega_m (t)= \omega_\mathrm{ge}(t)m + \frac{\alpha}{2}\left(m^2 - m\right),
\end{equation}
where the angular transition frequency between the two lowest-energy states is given by
\begin{equation}
    \omega_\mathrm{ge}(t) = \frac{\sqrt{8E_\mathrm{C}E_\mathrm{J}\left[\Phi_\mathrm{ext}(t)\right]} - E_\mathrm{C}}{\hbar},
\end{equation}
in terms of the time-dependent Josephson energy of the qubit $E_\mathrm{J}\left[\Phi_\mathrm{ext}(t)\right] = E_\mathrm{J}^\mathrm{max}|\cos\left[\Phi_\mathrm{ext}(t)/\Phi_0\right]|$. The external flux is defined for the preparation and the Otto cycle stages as
{\small
\begin{align}\label{eq:Phiext}
    \Phi_{\mathrm{ext}}(t) =
    \begin{cases}
        \Phi_{\mathrm{DC}}, & 0 \leq t < t_{\mathrm{A}}, \\
        \Phi_{\mathrm{DC}} + \Phi_{\mathrm{AC}} \sin^2\left[\omega (t - t_{\mathrm{A}})\right], & t_{\mathrm{A}} \leq t < t_{\mathrm{B}}, \\
        \Phi_{\mathrm{DC}} + \Phi_{\mathrm{AC}}, & t_{\mathrm{B}} \leq t < t_{\mathrm{C}}, \\
        \Phi_{\mathrm{DC}} + \Phi_{\mathrm{AC}} \sin^2\left[\omega (t -t_{\mathrm{C}}) + \frac{\pi}{2}\right], & t_{\mathrm{C}} \leq t <t_{\mathrm{D}}, \\
        \Phi_{\mathrm{DC}}, & \hspace{-7mm}t_{\mathrm{D}} \leq t <t_{\mathrm{D}} + \tau_4,
    \end{cases}
\end{align}}
where we use the notation for time instants and lengths defined in Fig.~\ref{fig:cycle_theory}b. Namely, we have $t_\mathrm{A} = \tau_\mathrm{p}$, $t_\mathrm{B}=\tau_\mathrm{p} + \tau_1$, $t_\mathrm{C} = \tau_\mathrm{p} + \tau_1 + \tau_2$, $t_\mathrm{D} = \tau_\mathrm{p} + \tau_1 + \tau_2 + \tau_3$, and we set $\tau_1=\tau_3=\pi/(2\omega)$. With these definitions, we know the change in the eigenenergies of the system, and we can compute the mean energies of the system from measured transmon populations.

The total transition rates in equations~\eqref{eq:dissip_emission} and \eqref{eq:dissip_absorption} are combinations of the intrinsic transition rates of the transmon and those induced by the QCR. For transmon eigenstates such that $E_m > E_n$, we obtain~\cite{teixeira_many-excitation_2024}
\begin{align}
    \Gamma_{mn}(t) &= \gamma_\mathrm{eg}^{(0)}(n + 1)(\overline{N}_{mn} + 1) + \Gamma_{mn}^\mathrm{QCR}(t),\nonumber\\
    \Gamma_{nm}(t) &= \gamma_\mathrm{eg}^{(0)}(n + 1)\overline{N}_{mn} + \Gamma_{nm}^\mathrm{QCR}(t),
\end{align}
where $\gamma_\mathrm{eg}^{(0)}$ is the intrinsic decay rate of the $\ket{\mathrm{e}}\leftrightarrow\ket{\mathrm{g}}$ transition, $\overline{N}_{mn}$ are the mean thermal occupation numbers of the transmon given by the Bose--Einstein distribution at the temperature of the intrinsic transmon bath as $\overline{N}_{mn} =~n_\mathrm{BE}(\omega_{mn},T_\mathrm{int}) = [\mathrm{e}^{\hbar\omega_{mn}/(k_\mathrm{B}T_\mathrm{int})} - 1]^{-1}$, and $\Gamma_{mn}^\mathrm{QCR}(t)$ are the QCR-induced transition rates. The intrinsic temperature $T_\mathrm{int}$ is obtained by fitting a Boltzmann distribution to the qubit populations at the thermal ground state. The transition rates from the QCR serve as the primary contribution to the total rates, especially for the first two states, and they can be approximated for $\omega_{mn}(t) =~\omega_{m}(t) -~\omega_{n}(t) > 0$ as~\cite{silveri_theory_2017,hsu_tunable_2020}
\begin{align}\label{eq:QCR-induced_transmon}
    \Gamma_{mn}^\mathrm{QCR}(t) &\approx \frac{\pi Z_\mathrm{aux}}{R_\mathrm{T}}\frac{g^2(n + 1)}{[\omega_{mn}(t) - \omega_\mathrm{aux}]^2}\nonumber\\
    &\times\sum_{\tau = \pm1} \Vec{F}[\tau eV_\mathrm{QCR}(t) + \hbar\omega_{mn}(t)],\nonumber\\
    \Gamma_{nm}^\mathrm{QCR}(t) &\approx \frac{\pi Z_\mathrm{aux}}{R_\mathrm{T}}\frac{g^2(n + 1)}{[\omega_{mn}(t) - \omega_\mathrm{aux}]^2}\nonumber\\
    &\times\sum_{\tau = \pm1} \Vec{F}[\tau eV_\mathrm{QCR}(t) - \hbar\omega_{mn}(t)],
\end{align}
where $Z_\mathrm{aux}$ and $\omega_\mathrm{aux}$ are the characteristic mode impe\-dance and resonance angular frequency of the auxiliary resonator, respectively, $g$ is its coupling strength with the transmon, $R_{\mathrm{T}}$ is the tunneling resistance of the NIS junction, and $\Vec{F}(E)$ is the normalized rate of forward quasiparticle tunneling controlled by the QCR voltage $V_{\mathrm{QCR}}(t)$~\cite{silveri_theory_2017}. The tunneling rates exhibit limited selectivity among the different transitions, though the low-lying transitions are somewhat more impacted than the high-lying transitions due to their higher transition frequencies. The effective temperature of the QCR can be obtained from the transition rates as~\cite{silveri_theory_2017}
\begin{equation}
    T_\mathrm{QCR}(t) = \frac{\hbar\omega_\mathrm{ge}(t)}{k_\mathrm{B}}\left[\mathrm{ln}\left(\frac{\Gamma_\mathrm{eg}^\mathrm{QCR}(t)}{\Gamma_\mathrm{ge}^\mathrm{QCR}(t)}\right)\right]^{-1},
\end{equation}
since the transition rates obey the detailed balance condition~\cite{silveri_theory_2017}
\begin{equation}
    \frac{\Gamma_\mathrm{eg}^\mathrm{QCR}(t)}{\Gamma_\mathrm{ge}^\mathrm{QCR}(t)} = \mathrm{e^{\hbar\omega_\mathrm{ge}(t)/[k_\mathrm{B}T_\mathrm{QCR}(t)]}}
\end{equation}
With these models, we can use the master equation in Eq.~\eqref{eq:master_eq} for analyzing as well as simulating the system dynamics.

\subsection*{Heat engine analysis}
For analyzing the characteristics of the heat engine based on the transmon-QCR system, we need to know the eigenfrequencies $\omega_m$ and the populations $p_n$ to estimate the terms of the mean energy
\begin{equation}
    \langle\hat{H}(t)\rangle = \sum_{m} \hbar\omega_m(t) p_m(t).
\end{equation}
Experimentally, we determine the energies by measuring the populations using single shots and the eigenfrequencies we obtain from two-tone spectroscopies. Theoretically, we solve the corresponding master equation and acquire the populations. For a simplified theoretical analysis, we concentrate below on an ideal quantum Otto cycle depicted in Fig.~\ref{fig:cycle_theory}a with a closed loop for which $W_\mathrm{i} + Q_\mathrm{c} + W_\mathrm{o} + Q_\mathrm{h} = 0$. In this ideal cycle, the eigenfrequencies stay constant for isochoric strokes and the populations stay constant in adiabatic strokes which yields the assumptions $\omega_m(t_\mathrm{A}) = \omega_m(t_\mathrm{D})$, $\omega_m(t_\mathrm{B}) = \omega_m(t_\mathrm{C})$, $p_m(t_\mathrm{A}) = p_m(t_\mathrm{B})$, and $p_m(t_\mathrm{C}) = p_m(t_\mathrm{D})$. Therefore, we can define the heat and work terms of the ideal cycle as 
\begin{align}
    W_\mathrm{o} &\equiv \langle\hat{H}\rangle(t_{\mathrm{B}})-\langle\hat{H}\rangle(t_{\mathrm{A}})\nonumber\\
    &=\sum_{m}\hbar\left[\omega_{m}(t_{\mathrm{B}})-\omega_{m}(t_{\mathrm{A}})\right]p_m(t_{\mathrm{A}}),\\
    Q_\mathrm{c} &\equiv \langle\hat{H}\rangle(t_{\mathrm{C}})-\langle\hat{H}\rangle(t_{\mathrm{B}})\nonumber\\
    &=\sum_{m}\hbar\omega_{m}(t_{\mathrm{B}})\left[p_m(t_{\mathrm{C}})-p_{m}(t_{\mathrm{A}})\right],\\
    W_\mathrm{i} &\equiv \langle\hat{H}\rangle(t_{\mathrm{D}})-\langle\hat{H}\rangle(t_{\mathrm{C}})\nonumber\\
    &=\sum_{m}\hbar\left[\omega_{m}(t_{\mathrm{A}})-\omega_{m}(t_{\mathrm{B}})\right]p_{m}(t_{\mathrm{C}}),\\
    Q_\mathrm{h} &\equiv \langle\hat{H}\rangle(t_{\mathrm{A}})-\langle\hat{H}\rangle(t_{\mathrm{D}})\nonumber\\
    &=\sum_{m}\hbar\omega_{m}(t_{\mathrm{A}})\left[p_{m}(t_{\mathrm{A}})-p_{m}(t_{\mathrm{C}})\right].
\end{align}
Here, the total work done by the driving field on the transmon is given by $W_\mathrm{tot} = W_\mathrm{i} + W_\mathrm{o}$, and the absorbed heat term from the heating stroke is $Q_\mathrm{abs} = Q_\mathrm{h} $. Thus, we have $W<0$ if the heat engine does positive work on the driving field. 

To further analyze the heat engine, we combine the heat and work terms with the time-dependent angular eigenfrequencies and obtain total work and absorbed heat expressions of the ideal cycle as
\begin{align}
     W_\mathrm{tot} \!&=\! -\hbar\omega_{\mathrm{ge}}(t_{\mathrm{A}})\!\left[1\!-\!\frac{\omega_{\mathrm{ge}}(t_{\mathrm{B}})}{\omega_{\mathrm{ge}}(t_{\mathrm{A}})}\right]\!\sum_{m} m\left[p_m(t_{\mathrm{A}})-p_m(t_{\mathrm{C}})\right]\!,\label{eq:totalW}\\
     Q_\mathrm{abs} &=\hbar\omega_{\mathrm{ge}}(t_{\mathrm{A}})\sum_{m} m\left[p_m(t_{\mathrm{A}})-p_m(t_{\mathrm{C}})\right]\nonumber\\
     &+\frac{\hbar\alpha}{2}\sum_{m}\left(m^2-m\right)\left[p_m(t_{\mathrm{A}})-p_m(t_{\mathrm{C}})\right].\label{eq:Q4}
\end{align}
Using the definitions of output power $P = -W_\mathrm{tot}/\tau_\mathrm{cyc}
$ and efficiency $\eta = -W_\mathrm{tot}/Q_\mathrm{abs}$, we observe that higher detuning during the adiabatic strokes leads to higher output power and efficiency as described by the $\omega_\mathrm{ge}(t_\mathrm{B})/\omega_\mathrm{ge}(t_\mathrm{A})$ term in the total work. In addition, the negative anharmonicity of the transmon increases the efficiency of the cycle since the absorbed heat is decreased in the heating stroke for more negative anharmonicities $\alpha$.

The ideal quantum Otto cycle analyzed above is challenging to realize experimentally owing to fluctuations arising from environmental noise and calibration challenges in perfectly closing the cycle. Consequently, the transmon populations slightly change during the adiabatic strokes, thus strictly speaking violating the assumptions for the ideal Otto cycle. Nevertheless, the violations can be made minor thanks to the exponential tunability of the QCR, and hence we compute the heat and work terms from the populations by numerically integrating the differential equations of heat and work from Eq.~\eqref{eq:internal_energy} of the internal energy as
\begin{align}
    W_{i\rightarrow j} &= \int_{i\rightarrow j}\delta W = \int_{i\rightarrow j}\sum_n \hbar p_n \mathrm{d}\omega_n,\\
    Q_{i\rightarrow j} &= \int_{i\rightarrow j}\delta Q = \int_{i\rightarrow j}\sum_n \hbar\omega_n \mathrm{d}p_n,
\end{align}
for stroke ($i\rightarrow\,j$). The numerical integration also acts as a filter for the fluctuations in the thermal state populations since the noise is partially averaged out during the numerical integration.

\subsection*{Numerical simulations}
To gain a theoretical prediction for the expected behavior of our experimentally realized QHE, we numerically simulate the dynamics of the heat engine using the master equation in Eq.~\eqref{eq:master_eq} for the preparation and Otto cycles. The pulses implementing the cycle are included in the time-dependent Hamiltonian of the qubit driven by an external flux for the flux tuning pulses, and in the Liouvillian dissipator operators steered by the QCR-induced transition rates $\Gamma_{nm}(t)$, which again depend on the bias voltage $V_\mathrm{QCR}(t)$. The characteristic quantities for the heat engine are calculated directly from the mean energies of the system obtained from the eigenfrequencies and the simulated populations. The master equation is simulated using the \emph{QuTiP} Python package for open-quantum-system simulations~\cite{johansson_qutip_2012} with system parameters acquired from the sample characterization and calibration.

\begin{figure}
    \centering
    \ifincludefigures\includegraphics[width=0.94\linewidth]{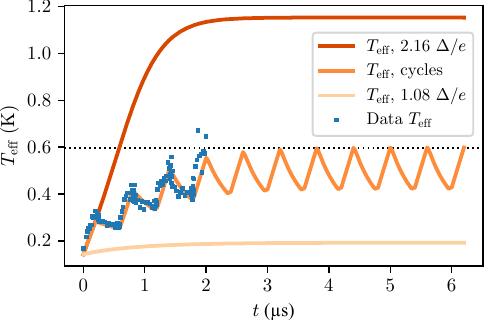}\fi
    \caption{\textbf{Simulation of the effective temperature of the transmon during a cyclic operation of the heat engine.} Simulated and measured effective temperatures of the transmon as functions of time for a bare heating pulse, bare cooling pulse, and Otto cycles with parameters corresponding to the experiments shown in Table~\ref{tab:parameters}. The dotted line displays the maximum simulated temperature for the Otto cycles.}
    \label{fig:ext_saturation_simulations}
\end{figure}

The simulations were used to acquire the transmon state populations during the Otto cycles that are compared with the experimentally obtained populations in Fig.~\ref{fig:main_results}a. Simulating a longer time evolution than that in Fig.~\ref{fig:main_results}a, but with identical system and pulse parameters, we further verify our claim that the cycle is nearly saturated already at slightly longer times than those used in the measurements. In Fig.~\ref{fig:ext_saturation_simulations}, we display the simulation results for the effective temperature of the transmon state for 10 cycles along with plain heating and cooling pulses. The figure shows that the state very accurately saturates after five cycles to a maximum temperature of 600 mK. Fitting exponential decay functions of form $\propto-\mathrm{exp(-t/\tau_\mathrm{sat})}$ to both the maximum and minimum points of the cycles yields a saturation time constant of value $\tau_\mathrm{sat} = 1.0\,\text{\textmu s}$. Thus, our measurements already span twice the length of the saturation time constant, further verifying that we are already measuring near the saturated effective temperatures.

From the temperatures during long heating and cooling pulses in Fig.~\ref{fig:ext_saturation_simulations}, we observe that during the cycles, the state saturates to temperatures slightly lower than the mean of those for the long heating and cooling pulses. The Carnot limit for the efficiency of the heat engine can be calculated from the temperatures of the thermal reservoirs by approximating them with those at the end of the long heating and cooling pulses, yielding a limit of $\eta_\mathrm{C} = 1 - T_\mathrm{c}/T_\mathrm{h}\approx0.83$ which greatly exceeds the measured efficiency. However, we can calculate the simulated efficiency of the cycle at the steady-state operation and obtain $\eta_\mathrm{steady} = 0.022$, matching the Otto efficiency obtained from the compression ratio. Therefore, the experimental steady-state operation of the Otto cycle, either after five cycles or with a tailored preparation heating pulse, should lead to the efficiency determined by the compression ratio.

\subsection*{Transmon readout and state identification}
Our qubit readout protocol relies on the well-established dispersive readout techniques with single shots, and the state identification is carried out using a multivariate Gaussian mixture model. As noted above, we obtain single-shot data by time-integrating the incident readout pulse after it has interacted with the qubit circuit. The readout pulse is of length 2\,\textmu s and the signal traces of the obtained quadrates I and Q are time-integrated from 0.4\,\textmu s to 1.0\,\textmu s. To enhance state identification, we optimize the readout frequency and power to obtain the best separation of single-shot point clouds in the IQ plane shown in Fig.~\ref{fig:cycle_calibration}c. 

For the single shots, we assume that the generated point clouds can be approximately characterized by a weighted sum of four two-dimensional Gaussian distributions that represent the lowest three eigenstates of the transmon $\{\ket{\mathrm{g}},\ket{\mathrm{e}},\ket{\mathrm{f}}\}$ in addition to the fourth distribution that accounts for several high-lying states, denoted by $\ket{\mathrm{h,i,j}}$. This weighted distribution, i.e., the Gaussian mixture model, can be expressed as 
\begin{align}
    G(\mathbf{x}) &=\!\!\!\!\sum_{m=\ket{\mathrm{g}},\ket{\mathrm{e}},\ket{\mathrm{f}},\ket{\mathrm{h,i,j}}}\!\!\!\!w_m\mathcal{N}(\mathbf{x},\boldsymbol{\mu}_m,\Sigma_m),\nonumber\\
    &=\!\!\!\!\sum_{m=\ket{\mathrm{g}},\ket{\mathrm{e}},\ket{\mathrm{f}},\ket{\mathrm{h,i,j}}}\!\!\!\!w_m \frac{1}{\sqrt{2\pi}|\Sigma_m|^{1/2}}\mathrm{e}^{-\frac{1}{2}(\mathbf{x} - \boldsymbol{\mu}_m)^T\Sigma_m^{-1}(\mathbf{x} - \boldsymbol{\mu}_m)},
\end{align}
where $\mathbf{x}$ is a two-dimensional vector in the IQ plane, $\boldsymbol{\mu}_m$ is the mean vector of the Gaussian component $m$, $\Sigma_m$ is the covariance matrix, and $w_m$ is the corresponding weight of the Gaussian distribution. The fit of the single-shot data for this GMM is optimized by an expectation-maximization algorithm, which is an iterative scheme for machine-learning parameters through maximum likelihood estimates~\cite{deisenroth_mathematics_2020}. In practice, the optimization is implemented by using the \emph{scikit-learn} Python library~\cite{pedregosa_scikit-learn_2011} to obtain the optimal Gaussian distributions for the single-shot point clouds. The model is trained on a calibration data set combining 10\,000 single shots for each prepared state of the lowest three eigenstates. 

Using the calibrated GMM distributions shown in Fig.~\ref{fig:cycle_calibration}c, we estimate the populations of the transmon states in the single-shot data acquired in the Otto cycle measurements. For each point in the measurement data in Figs.~\ref{fig:main_results}a and \ref{fig:main_results}b, we measure 10\,000 single shots and calculate the uncorrected populations of the states using the estimated probability equation
\begin{equation}\label{eq:prob1}
    \Tilde{p}_m = \frac{\tilde{N}_m}{\tilde{N}_{\ket{\mathrm{g}}} + \tilde{N}_{\ket{\mathrm{e}}} + \tilde{N}_{\ket{\mathrm{f}}} + \tilde{N}_{\ket{\mathrm{h,i,j}}}},
\end{equation}
where $\tilde{N}_m$ is the uncorrected number of single-shot data points inside the $a\boldsymbol{\sigma}_m$-boundary ellipse corresponding to state $\ket{m}$ where $\boldsymbol{\sigma}_m$ is the standard deviation vector for the $x$ and $y$ components of the Gaussian distribution and $a$ is the scaling coefficient for the size of the ellipse. Essentially, we count the number of data points from the single-shot measurements that are inside the $a\boldsymbol{\sigma}_m$-ellipse centered at the mean value $\boldsymbol{\mu}_m$ for each state ellipse and normalize them using the total number of points inside these ellipses. The scaling coefficient $a=0.4$ was employed to mitigate false counts from incorrect states within each ellipse while maintaining a sufficient number of data points to characterize each state accurately. 

To further account for the overlap of GMM distributions, we employed a correction matrix on the measured populations generated by sampling 10 million random points from each of the obtained 2D Gaussian distributions and then calculating the normalized number of points lying inside each boundary ellipse representing each state. The resulting correction matrix is 
{\footnotesize
\begin{equation}
    M=\begin{bmatrix}
       0.3934 & 0.0350 & \num{9.200e-6} & 0.0000\\
       0.0309 & 0.3936 & 0.0314 & \num{8.500e-6}\\
       \num{6.770e-5} & 0.0281 & 0.3934 & 0.0364\\
       \num{7.000e-6} & \num{6.777e-4} & 0.0370 & 0.3935
    \end{bmatrix},
\end{equation}}
where the matrix elements $M_{i,j}$ are the ratios of points counted in the boundary ellipse of distribution $j$ over points sampled from the distribution $i$. Thus, to obtain the corrected populations of the states $p_m$, we do not apply Eq.~\eqref{eq:prob1} for the uncorrected counts $\tilde{N}_m$ but instead for the corrected counts $N_m$ as
\begin{equation}
    \begin{bmatrix}
        N_{\ket{\mathrm{g}}} \\ 
        N_{\ket{\mathrm{e}}} \\ 
        N_{\ket{\mathrm{f}}} \\
        N_{\ket{\mathrm{h,i,j}}}
    \end{bmatrix} =   M^{-1}\begin{bmatrix}
        \tilde{N}_{\ket{\mathrm{g}}} \\ 
        \tilde{N}_{\ket{\mathrm{e}}} \\ 
        \tilde{N}_{\ket{\mathrm{f}}} \\
        \tilde{N}_{\ket{\mathrm{h,i,j}}}
    \end{bmatrix}.
\end{equation}

From the obtained populations, we estimate the effective temperature $T_\textrm{eff}$ of the qubit by finding the thermal state that best matches these populations.  Specifically, we maximize the fidelity $F(T_\mathrm{eff}) = \left(\sum_m \sqrt{p_m^\mathrm{th}(T_\mathrm{eff})p_m^\mathrm{meas}} \right)^2$, which quantifies the similarity between the measured populations $p_m^\mathrm{meas}$ and the thermal state populations $p_m^\mathrm{th}$ of state $m$, computed from the qubit eigenfrequencies $\omega_m$ at temperature $T_\mathrm{eff}$. 

To account for statistical uncertainty in the measured populations, we carry out Monte Carlo resampling. For each dataset, we draw 1000 samples from the multinomial distribution defined by $p_m^\mathrm{meas}$. For each sample, we again determine $T_\mathrm{eff}$ by maximizing the fidelity with respect to temperature. The resulting distribution yields the mean and standard deviation of the effective temperature for each set of measured populations.

\bibliographystyle{naturemag}
\bibliography{biblio}

\section*{Acknowledgments}
We thank Aashish Sah, Mikko Tuokkola, András M. Gunyhó, Heikki Suominen, Bayan Karimi, Jukka Pekola, Ilari Mäkinen, Christoforus Satrya, Yoshiki Sunada, Suman Kundu, and Joachim Ankerhold for discussions.

\subsection*{Funding}
This work was funded by the Research Council of Finland Centre of Excellence program (project Nos. 352925 and 336810) and grant Nos. 316619 and 349594 (THEPOW). We also acknowledge funding from the European Research Council under Advanced Grant No. 101053801 (ConceptQ), personal funding from the Finnish Cultural Foundation (T.U.), Jane and Aatos Erkko Foundation through the project SystemQ, and the provision of facilities and technical support by Aalto University at OtaNano--Micronova Nanofabrication Centre.

\subsection*{Author contributions}
T.U. developed the pulsing protocols for the measurements, carried out the data analysis, and wrote the manuscript with input from all the authors. T.U. and T.M. conducted the experiments. T.M. fabricated the sample. T.M. and W.T. contributed to developing the devices and the measurement scheme. T.U., W.T., and T.M. contributed to the data analysis scripts. W.T., M.R., and M.M. gave theory support. M.M. supervised the work in all respects. 

\subsection*{Competing interests}
M.M. declares that he is a Co-Founder and Shareholder of quantum companies IQM Finland Oy and QMill Oy. The other authors declare no competing interests. 

\subsection*{Data and code availability}
The data and special codes that support the findings of this study are available at \href{https://doi.org/10.5281/zenodo.14935889}{https://doi.org/10.5281/zenodo.14935889}.

\end{document}